\documentclass[aps,pra,a4paper,superscriptaddress,longbibliography,reprint]{revtex4-1}

\usepackage[T1]{fontenc}
\usepackage{newtxmath,newtxtext}

\usepackage{amsmath}
\usepackage{amsfonts}
\usepackage{bm}
\usepackage{bbm}

\usepackage{graphicx}
\usepackage{color}

\newcommand{\fig}[1]{Fig.~\ref{#1}}

\begin{document}

\title{Anomalous Enhancement of the Electrocatalytic Hydrogen
Evolution Reaction in AuPt Nanoclusters}

\author{Jiahui Kang}
\affiliation{Department of Applied Physics,
Aalto University, 02150, Espoo, Finland}

\author{Jan Kloppenburg}
\affiliation{Department of Chemistry and Materials Science,
Aalto University, 02150, Espoo, Finland}

\author{Jiali Sheng}
\affiliation{Department of Applied Physics,
Aalto University, 02150, Espoo, Finland}

\author{Zhenyu Xu}
\affiliation{Department of Applied Physics,
Aalto University, 02150, Espoo, Finland}

\author{Kristoffer Meinander}
\affiliation{Department of Bioproducts and Biosystems,
Aalto University, 02150, Espoo, Finland}

\author{Hua Jiang}
\affiliation{Department of Applied Physics,
Aalto University, 02150, Espoo, Finland}

\author{Zhong-Peng Lv}
\affiliation{Department of Applied Physics,
Aalto University, 02150, Espoo, Finland}

\author{Esko I. Kauppinen}
\affiliation{Department of Applied Physics,
Aalto University, 02150, Espoo, Finland}

\author{Qiang Zhang}
\affiliation{Department of Applied Physics,
Aalto University, 02150, Espoo, Finland}

\author{Xi Chen}
\affiliation{Department of Applied Physics,
Aalto University, 02150, Espoo, Finland}
\affiliation{School of Physical Science and Technology, Lanzhou University,
Lanzhou, Gansu 730000, China}
\affiliation{Lanzhou Center for Theoretical Physics and Key Laboratory for
Quantum Theory and Applications of the Ministry of Education, Lanzhou University,
Lanzhou, Gansu 730000, China}

\author{Olli Ikkala}
\affiliation{Department of Applied Physics,
Aalto University, 02150, Espoo, Finland}

\author{Miguel A. Caro}
\email{mcaroba@gmail.com}
\affiliation{Department of Chemistry and Materials Science,
Aalto University, 02150, Espoo, Finland}

\author{Bo Peng}
\email{pengbo006@gmail.com}
\affiliation{Department of Applied Physics,
Aalto University, 02150, Espoo, Finland}
\affiliation{Department of Materials Science, Advanced Coating Research Center of Ministry of Education of China, Fudan University, Shanghai 200433, China}

\date{\today}

\begin{abstract}
\begin{center}
\textbf{Abstract}
\end{center}

Energy- and resource-efficient electrocatalytic water splitting is of paramount importance to
enable sustainable hydrogen production. The best bulk catalyst for the hydrogen
evolution reaction (HER), i.e., platinum, is one of the scarcest elements on Earth. The
use of raw material for HER can be dramatically reduced by utilizing nanoclusters.
In addition, nanoalloying can further improve the performance of these nanoclusters.
In this paper, we present results for HER on nanometer-sized ligand-free AuPt
nanoclusters grafted on carbon nanotubes. These results demonstrate excellent
monodispersity and a significant reduction of the overpotential for the electrocatalytic
HER. We utilize atomistic machine learning techniques to elucidate the atomic-scale
origin of the synergistic effect between Pt and Au. We show that the presence of
surface Au atoms, known to be poor HER catalysts, in a Pt(core)/AuPt(shell)
nanocluster structure, drives an anomalous enhancement of the inherently high
catalytic activity of Pt atoms. 
\end{abstract}

\maketitle

\section{Introduction}

With the expansion of industry and the growth of the world population,
the demand for energy from non-renewable fossil fuels has increased significantly,
inevitably generating large amounts of carbon emissions and thus causing climate
change and energy scarcity~\cite{seh_2017,browne_2019,hu_2019,li_2021}. Hydrogen
is considered as one of the most essential clean and renewable energy sources for
alternative high-efficiency energy conversion and storage technologies, especially
when coupled to green electricity from such renewable resources as wind, tidal,
and solar~\cite{han_2016}. Water is the most readily accessible source of hydrogen.
Thus, energy-efficient electrocatalytic water splitting is the most attractive way
to produce hydrogen. The hydrogen evolution reaction (HER) is a key half-reaction of
water splitting on the cathode, which involves multi-step reactions, hindering the
efficiency of water splitting. Therefore, the availability of high-efficiency catalysts
for HER is critical for accelerating the sluggish reaction and decreasing the dynamic
overpotential~\cite{zhu_2018,zhang_2016,zou_2015,han_2016}. Currently, noble
metal-based materials are the most widely used to decrease the overpotential, such as
Pt-based catalysts for HER~\cite{jiang_2019,zhang_2018,sanati_2022}. Due to their cost
and scarcity, reducing the amount of raw material is of paramount importance when utilizing Pt and
other noble metals. For this reason, replacing bulk catalysts with their nanostructured
forms is a widely explored strategy. Pt-based nanoparticles have been studied for many
years as a means to decrease the usage of raw Pt and improve the efficiency and stability
of Pt-based catalysts for HER~\cite{jiang_2019,zhang_2018}.

Thanks to the small size and large specific area of nanoclusters (NCs), which involve
monodipersity, NCs are expected to have more active catalytic sites than bulkier
nanoparticles, making them a promising nanomaterial for water splitting catalysis. Moreover,
compared with nanoparticles, the electronic structure and properties of NCs are more strongly
dependent on the applied potential~\cite{jin_2015,chen_2015,zeng_2015}. The size, morphology
and composition of NCs are the three key factors affecting their
properties~\cite{gilroy_2016,wang_2015,sankar_2012,xia_2009,li_2010,zhou_2012}.
With regard to composition, alloyed NCs are designed to improve the physicochemical
activity because of the possible synergistic effects in multi-metallic NCs, where the
properties of the alloy differ from the properties of the mono-metallic
constituents~\cite{gilroy_2016,wang_2015,yao_2018,chen_2016,xu_2018,wang_2018}.
For instance, Pt/Pd-doped MAu(SR)$_{18}$ gold NCs, where R is the ligand,
were extensively studied in recent
years because of the interesting alloying effect that regulates the geometry,
structure and physicochemical properties, including thermal stability, reactivity,
catalytic performance, and magnetism~\cite{qian_2012,xie_2012,fields-zinna_2009}.

In this paper, we show that alloying Pt NCs, which are intrinsically catalytically
active towards HER~\cite{jiang_2019,zhang_2018}, with Au, which is catalytically
inactive for the same reaction, allows the formation of bimetallic  NCs with
superior and tunable catalytic activity. 

Another important consideration is the tradeoff between NC activity and stability,
because in industrial-scale applications the active materials need to remain
catalytically active for extended periods of time to be economically viable.
Ligand-protected NCs are stable but insufficiently efficient for catalysis. Removing
the catalytically inert ligand of NCs by annealing is an effective way to improve
their performance~\cite{liu_2016}. However, the NCs can aggregate after ligand
removal and suffer from instability during the electrocatalytic process. Grafting the
NCs to a stable support, e.g., conductive carbon, is a widely accepted way to prevent
the aggregation of ligand-free NCs by immobilizing them on the support~\cite{shao_2006}.
Carbon nanotubes (CNTs) are an attractive support for electrocatalysis because of their
unique structure and intrinsic properties such as high surface area, high chemical stability,
high electrical conductivity, and insolubility in most solvents~\cite{liu_2009,nassr_2014}.
In this work, we precisely regulate the catalytic activity by optimizing the Au:Pt ratio
of Pt (core)-AuPt (shell) NCs to increase the number of active sites and enhance the
intrinsic activity towards HER. The NCs are grafted to CNT substrates to facilitate the
electric conductivity and prevent agglomeration. The ligand-free  Au$_{0.4}$Pt$_{0.6}$ NC-CNT-A (where
A stands for annealed) complex shows the highest HER activity among all studied NC-CNT
complexes, with an overpotential ($\eta$) of 27~mV at 10~mA~cm$^{-2}$, a high mass activity of
7.49~A~mg$^{-1}$ at $\eta = 100$~mV, and a high turnover frequency (TOF) of 7.63~s$^{-1}$, outperforming the
majority of reported Pt-based catalysts to date. 

To provide a mechanistic understanding of the origin of this anomalous enhancement,
we resort to combining the experimental results with atomistic simulation~\cite{seh_2017}.
We carry out simulated-annealing molecular dynamics (MD) simulations with a custom-made
machine learning potential (MLP) for AuPt:H to elucidate the nature of the ligand-free AuPt NC
structure. This analysis reveals a strong surface segregation of Au atoms and consequent
core (Pt)-shell (AuPt) structure of the annealed AuPt NCs. We also perform heuristic
Markov-chain grand-canonical Monte Carlo (GCMC) simulations at variable (electro) chemical
potential. These suggest that, despite their low intrinsic activity, the presence of Au atoms
drives an enhancement of H adsorption on neighboring Pt atoms. In addition to providing a
mechanistic understanding of the H-adsorption process and its interplay with the AuPt
alloy nanostructure, these simulations predict, in full agreement with the experimental
observations, that H adsorption is more favorable on AuPt at a low Au fraction than in
pure Pt, resulting in excellent catalytic performance for HER. This enhanced activity stems
from the ability of the exposed adsorption Pt sites on the NC surface to bind hydrogen at
higher electrode potential in the presence of neighboring Au atoms. This cooperative
enhancement is ``anomalous'' in the sense that it follows the opposite trend that would be
expected from interpolating between the behavior of pure ligand-free Pt NCs (excellent HER catalysts)
and pure ligand-free Au NCs (poor HER catalysts). 

\section{Experimental section}

\subsection{Preparation of AuPt nanoclusters and AuPt-CNT-A}

Chemicals and materials: L-glutathione (GSH) in the reduced form ($\ge 98.0$~\%, Sigma-Aldrich),
sulfuric acid (puriss., Sigma, 95-97~\%), chloroplatinic acid hexahydrate
(H$_2$PtCl$_6$·6H$_2$O, ACS, $\ge 37.50$~\% Pt basis, Sigma), gold(III) chloride trihydrate
(HAuCl$_4$·3H$_2$O, $\ge 99.9$~\% trace metals basis, Sigma), dialysis membrane (3.5~kD).

Synthesis of the (ligand-protected) Au$_{0.4}$Pt$_{0.6}$ NCs:
The NCs were prepared following the protocol of Ref.~\cite{luo_2012} for pure Au NCs 
but adjusting the ratio of Au and Pt precursors
according to the desired final NC stoichiometry
(0.4/0.6 feed ratio for Au$_{0.4}$Pt$_{0.6}$). The NCs with other compositions had
their feed ratios modified accordingly.

Synthesis of the (ligand-protected) Au$_{0.4}$Pt$_{0.6}$-CNT film: The H$_2$SO$_4$-processed CNT films were
put into 2~mL Au$_{0.4}$Pt$_{0.6}$ NCs dispersion for 20~h.

Synthesis of (ligand-free) Au$_{0.4}$Pt$_{0.6}$-CNT-A film: An Au$_{0.4}$Pt$_{0.6}$-CNT film was put into
a tube furnace at 300~$^\circ$C for 30~min annealing under N$_2$ protection. After annealing,
the Au$_{0.4}$Pt$_{0.6}$-CNT-A film was prepared.

\subsection{Structural characterization of AuPt nanoclusters and AuPt-CNT-A}

The transmission electron microscopy (TEM) characterizations and energy dispersive X-ray
spectroscopy (EDS) mapping were performed on a JEOL JEM-2200FS operated at 200~kV. All
particle size distributions are calculated based on a count of $\sim 100$ particles.
The X-ray diffraction (XRD) patterns were recorded using the Rigaku Smart Lab X-ray
diffractometer with Cu-K$\alpha$ radiation. The small angle X-ray scattering (SAXS) and
wide-angle X-ray scattering (WAXS) measurements were conducted by Xenocs Xeuss 3.0.
The surface component measurements were performed with a Kratos AXIS Ultra DLD X-ray
photoelectron spectrometer (XPS) using a monochromated Al-K$\alpha$ X-ray source (1486.7~eV)
run at 100~W. All CNT spectra were charge-corrected relative to the position of the graphitic
carbon at a binding energy of 284.2~eV. XPS of the ligand-protected
AuPt NCs without CNT is measured after that the solutions
were freeze-dried to powders. All powder spectra were also charge-corrected relative to the
position of C-C bonding at a binding energy of 284.8~eV. The inductively coupled
plasma-optical emission spectroscopy (ICP-OES) test was performed using an Agilent 730 series
ICP optical emission spectrometer to determine the actual Au and Pt element contents.
Fourier-transform infrared spectroscopy (FTIR) was performed with a PerkinElmer FTIR
with attenuated total reflectance (ATR). The X-ray absorption near-edge structure (XANES)
was performed in transmission mode at the MAX IV Balder beamline, Sweden.

\subsection{Electrochemical measurements}

HER measurements were performed on a Metrohm Autolab electrochemical workstation at room
temperature. CNT films with dimensions of $1 \times 0.2$~cm$^2$ were used as the working
electrode. Ag/AgCl (3~M KCl) and graphite rods were used as the reference and the counter
electrodes (RE and CE), respectively. The Ag/AgCl electrode was calibrated with respect to a reversible
hydrogen electrode (RHE). The sample surface area for electrochemical test is 0.2~cm$^2$.
The HER measurement was carried out in an N$_2$-saturated 0.5~M H$_2$SO$_4$ aqueous solution.
Linear sweep voltammetry (LSV) was conducted at a scan rate of 5~mV~s$^{-1}$. The stability
test was performed at a current density of 10~mA cm$^{-2}$ for 24~h within the N$_2$-saturated 0.5~M
H$_2$SO$_4$ aqueous solution. The electrochemical surface area (ECSA) measurements were
conducted at a potential between 0.2 and 0.4~V versus RHE. Electrochemical impedance
spectroscopy (EIS) measurements were done from 100~kHz to 0.1~Hz. The LSV plots and chronoamperometric measurement are with IR compensation to eliminate the resistance overpotential. The IR
compensation can be expressed by the following equation:
$E_\text{corrected} = E_\text{uncorrected} - I \times R$~\cite{shi_2016}.

\section{Results and discussion}

\subsection{Synthesis and morphologies of AuPt-CNT-A}

\begin{figure*}[p]
    \centering
    \includegraphics[width=0.85\linewidth]{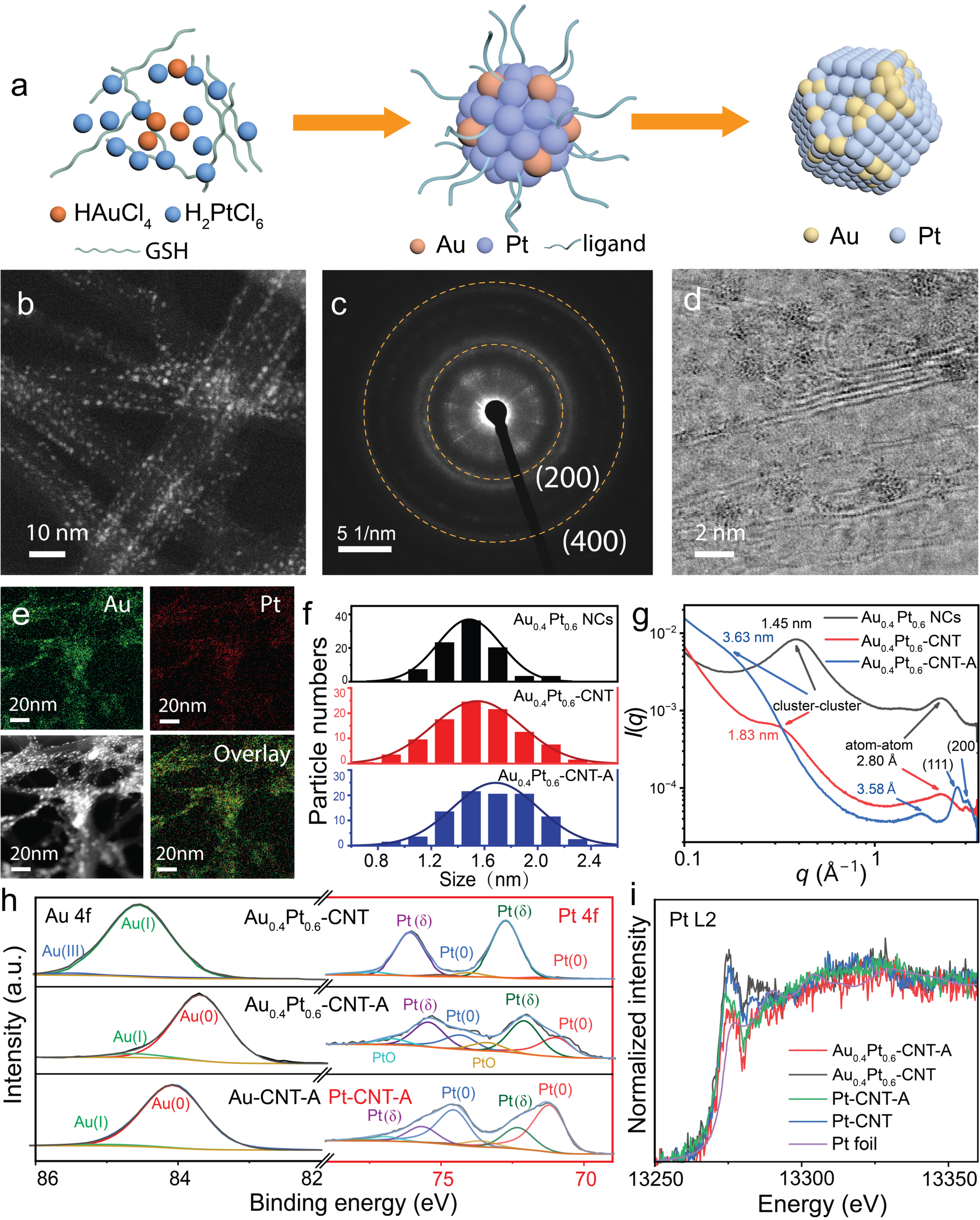}
    \caption{a) Schematics of the preparation of AuPt NCs on CNT with annealing.
    b) High-angle annular dark-field scanning transmission electron microscopy (HAADF-STEM) image of Au$_{0.4}$Pt$_{0.6}$-CNT-A, and the corresponding c) Selected area electron diffraction (SAED)
    patterns of Au$_{0.4}$Pt$_{0.6}$ nanoclusters on CNT with annealing.
    d) High-angle annular bright-field scanning transmission electron microscopy (HAABF-STEM) image of Au$_{0.4}$Pt$_{0.6}$-CNT-A.
    e) STEM-EDS (energy dispersive X-ray spectroscopy) elemental mapping of
    Au$_{0.4}$Pt$_{0.6}$-CNT-A. f) Corresponding particle size statistics and
    g) wide-angle X-ray scattering (WAXS) results of Au$_{0.4}$Pt$_{0.6}$ NCs,
    Au$_{0.4}$Pt$_{0.6}$-CNT, and Au$_{0.4}$Pt$_{0.6}$-CNT-A. h) Au 4f and Pt 4f
    XPS of the grafted Au$_{0.4}$Pt$_{0.6}$-CNT, and Au$_{0.4}$Pt$_{0.6}$-CNT-A film,
    Au 4f XPS spectra of pure Au-CNT-A film, and Pt 4f XPS spectra of with
    pure Pt-CNT-A film. i) Normalized X-ray absorption near-edge structure (XANES)
    spectra at Pt L$_2$-edge of Au$_{0.4}$Pt$_{0.6}$-CNT-A and Au$_{0.4}$Pt$_{0.6}$-CNT,
    Pt-CNT-A, Pt-CNT and Pt foil.}
    \label{01}
\end{figure*}

The ligand-free AuPt-CNT-A was synthesized by the following steps: 1) synthesis of the ligand-protected AuPt nanoclusters;
2) ligand-protected AuPt NCs grafted onto the CNTs and 3) annealing, as schematically shown in \fig{01}a.
To elucidate the morphology of the CNTs before and after processing, transmission electron
microscopy (TEM) characterization was performed. The characteristics of the pristine CNTs
(before processing) are summarized in Fig.~S1. The pristine CNTs are grown by nucleation on Fe nanoparticles, formed from the decomposition of the ferrocene precursor. These Fe nanoparticles are encapsulated by
uniform graphitic carbon layers (Fig.~S1c), with interlayer spacing
$d(002) = 0.32$-$0.36$~nm~\cite{park_2018}. We further identify spacings of 0.206~nm
of the particles could originate from the (110) plane of bcc Fe~\cite{edalati_2022} or the (102) plane
of Fe$_3$C~\cite{huang_2020}, respectively. After washing with H$_2$SO$_4$, the sulfated
CNTs (Fig.~S2c) retain their structure. The XRD patterns (Fig.~S4) of the Au$_{0.4}$Pt$_{0.6}$-CNT and
Au$_{0.4}$Pt$_{0.6}$-CNT-A samples show a series of broad Bragg peaks, suggesting low
crystallinity of the of the grafted NCs and annealed NCs on CNT. 

On the other hand, the TEM analysis shows a certain degree of crystallization of the ligand-free Au$_{0.4}$Pt$_{0.6}$ NCs on CNT 
 film with annealing, as shown in \fig{01}b-d.
The selected area electron diffraction (SAED) (\fig{01}c) demonstrates diffraction
bright rings at d-spacings 2.04~\AA{} and 1.02~\AA{},
respectively, corresponding to the (200) and (400) plane of AuPt~\cite{xu_2008,feng_2017},
with other  rings originated from the (110), (100), and (002) graphitic carbon, which
is similar to the processed CNTs (Figure S3).
The broad peaks in the XRD pattern are likely a consequence of the small size of the
NCs~\cite{zhang_2003}. The scanning transmission electron microscopy coupled
energy-dispersive X-ray spectroscopy (STEM-EDS) element mapping in \fig{01}e shows the
distribution of Au and Pt on the NCs across the selected area. For comparison, as shown
in Fig.~S5, the (ligand-protected) Au$_{0.4}$Pt$_{0.6}$ NCs as precursors are monodisperse with an average
diameter of 1.5~nm (polydispersity index (PdI) = 0.025), suggesting the ligand stabilizes
the structure of the NCs~\cite{luo_2012}. The (ligand-protected) Au$_{0.4}$Pt$_{0.6}$ NCs
are amorphous, which can be confirmed by their SAED patterns (Fig.~S5b). The EDS elemental
mapping in Fig.~S5d-h further confirms the homogeneous distribution of both Au and Pt
elements in Au$_{0.4}$Pt$_{0.6}$ NCs. After grafting, the (ligand-protected) Au$_{0.4}$Pt$_{0.6}$
NCs are monodispersed on the CNTs ($\sim 1.54$~nm, PdI = 0.047), as shown in Fig.~S6.
Furthermore, after annealing, as shown in Fig.~S7, the (ligand-free) Au$_{0.4}$Pt$_{0.6}$ NCs ($\sim 1.68$~nm,
PdI = 0.036) are still monodispersely distributed over the CNT matrix with a narrow size
distribution (\fig{01}f), and the sizes are almost unchanged in comparison to the (ligand-protected) Au$_{0.4}$Pt$_{0.6}$ NCs on CNT prior to annealing.

\subsection{WAXS and SAXS characterizations}

In order to further obtain the structural features (size and shape) of the NCs on CNTs,
small angle X-ray scattering (SAXS) and wide angle X-ray scattering (WAXS) experiments were
performed on the (ligand-free) Au$_{0.4}$Pt$_{0.6}$-CNT-A film, (ligand-protected) Au$_{0.4}$Pt$_{0.6}$-CNT film, CNT film
and (ligand-protected) Au$_{0.4}$Pt$_{0.6}$ NCs (Figs.~\ref{01}g, S8 and S9, and Table~S1). A log-normal
distribution of diameters was used to fit the WAXS data. The average diameters $\langle d \rangle$
of the NCs obtained thereof are shown in Table~S1. 
The value of 2.80~\AA{} in Au$_{0.4}$Pt$_{0.6}$-CNT and Au$_{0.4}$Pt$_{0.6}$ NCs shows the
typical distance between Au and Pt atoms in an NC. The distances of 2.33~\AA{} and
2.04~\AA{} in Au$_{0.4}$Pt$_{0.6}$-CNT-A can be assigned to the lattice spacings of the
(111) and (200) planes of AuPt, respectively, which indicate a certain degree of AuPt
crystallization after annealing, which is consistent with TEM~\cite{xu_2008,feng_2017}.
It is worth noting that the distance of 3.58~\AA{} in Au$_{0.4}$Pt$_{0.6}$-CNT-A corresponds
to the interlayer spacing in graphitic carbon. This peak becomes apparent only after
annealing, possibly because of the shift (and splitting) of the partly overlapping AuPt
peak. From the Kratky plot (Figure S9a),
Au$_{0.4}$Pt$_{0.6}$-CNT shows a peak around 1.83~nm, suggesting that this is the
average distance between AuPt NCs distributed on the CNTs. After removing the ligand,
Au$_{0.4}$Pt$_{0.6}$-CNT-A shows 3.63~nm for the distance between the NCs, which suggests
AuPt NCs migration on the CNTs during annealing. The SAXS data in Fig.~S9b indicates the
average separation between CNTs in the CNT bundles is expanded from 33~nm to >50~nm after
deposition of the NCs.

\subsection{Atomistic structure and XPS characterization}

To determine the atomistic structure of Au$_{0.4}$Pt$_{0.6}$-CNT-A, the chemical
composition was further examined by X-ray photoelectron spectroscopy (XPS) analysis.
The XPS of CNT films as the substrate is shown in Fig.~S10. The survey spectra of (ligand-protected)
Au$_{0.4}$Pt$_{0.6}$ NCs, (ligand-protected) Au$_{0.4}$Pt$_{0.6}$-CNT, and (ligand-free) Au$_{0.4}$Pt$_{0.6}$-CNT-A
shown in Fig.~S11 confirm the presence of Au, Pt, C, N, S, and O elements and similar
chemical composition of the Au$_{0.4}$Pt$_{0.6}$ NCs and the grafted nanoclusters
(Au$_{0.4}$Pt$_{0.6}$-CNT). The normalization of XPS peaks by their relative sensitivity
factors yielded the average composition of Au$_{0.4}$Pt$_{0.6}$ NCs, Au$_{0.4}$Pt$_{0.6}$-CNT,
and Au$_{0.4}$Pt$_{0.6}$-CNT-A and is shown in Table~S2. The increased C/O ratio after
annealing suggests the decrease of OH groups in Au$_{0.4}$Pt$_{0.6}$-CNT-A.  

\subsection{XPS characterization of the pristine ligand-protected Au$_{0.4}$Pt$_{0.6}$ NCs}

The Au 4f$_{7/2}$ pattern of the Au$_{0.4}$Pt$_{0.6}$ NCs (Fig.~S12a and Table~S3) can
be deconvoluted to two peaks at 85.0~eV and 84.1~eV, which can be assigned to Au(III)
and Au(I), respectively~\cite{luo_2012,duan_2016}. The main peak of Pt 4f$_{7/2}$ of the pristine
Au$_{0.4}$Pt$_{0.6}$ NCs (Fig.~S12b) is at 72.2~eV, which belongs to Pt($\delta +$)
($0<\delta<2$)~\cite{dablemont_2008}. Also a small amount of Pt(0) was found in pristine
Au$_{0.4}$Pt$_{0.6}$ NCs~\cite{imaoka_2017}. Pristine Au$_{0.4}$Pt$_{0.6}$ NCs with a
high content of S for thiol (Fig.~S11e, Tables~S4 and S5) hints towards the formation
of thiolate complexes~\cite{luo_2012}. The quantitative analysis indicates that,
for pristine Au$_{0.4}$Pt$_{0.6}$ NCs, $\sim 20$~\% of the sulphur species with the
dominant S 2p$_{3/2}$ peak at 166.4~eV are oxidized S 2p components. The S 2p core level
is characterized by two S 2p$_{3/2}$ and S 2p$_{1/2}$ spin-orbit split doublets. The
dominant S 2p$_{3/2}$ peak at 162.7~eV and accompanying S 2p$_{1/2}$ peak at 163.8~eV
can be assigned to thiolate groups, confirming the formation of AuPt-S thiolate bonds
in pristine Au$_{0.4}$Pt$_{0.6}$ NCs. According to previous research, ligand-protected
Au NCs present a core-shell nanostructure with a Au(I)-thiolate complex as the shell and
Au(0) as the core~\cite{luo_2012}. Based on these results and our XPS analysis, we
speculate that the (ligand-protected) pristine Au$_{0.4}$Pt$_{0.6}$ NCs have a similar
core-shell nanostructure, i.e., Au(I)/Pt($\delta +$)-thiolate complex as the shell
and Pt(0) as the core.

\subsection{XPS characterization of grafted ligand-protected Au$_{0.4}$Pt$_{0.6}$-CNT}

To further determine the location of Pt atoms of the NCs after being grafted onto the CNTs,
the Au 4f$_{7/2}$ and Pt 4f$_{7/2}$ XPS spectra of the grafted Au$_{0.4}$Pt$_{0.6}$-CNT
sample were measured. Again, the Pt 4f$_{7/2}$ spectrum of grafted Au$_{0.4}$Pt$_{0.6}$-CNT
can be deconvoluted into two peaks at 72.7~eV and 71.5~eV, where the latter can be assigned
to Pt(0)~\cite{imaoka_2017} (\fig{01}h). Additionally, the peak position at 72.7~eV is at
much lower energy than the reference Pt(II) peak, from which the 2+ valence state of Pt
is expected to appear with a positive shift of +2.4~eV with respect to the Pt(0) peak.
Therefore, it is reasonable to assign the peak at 72.7~eV to a state with intermediate
valency Pt($\delta +$) ($0<\delta<2$)~\cite{dablemont_2008}. Besides, the grafted
Au$_{0.4}$Pt$_{0.6}$-CNT shows a high content of S, attributed to the thiol groups
from the surface complex, which also confirms that the Au$_{0.4}$Pt$_{0.6}$ NCs on CNT
are thiol-coated~\cite{dablemont_2008}. Moreover, the Au 4f peak at 84.5~eV exhibited a
positive shift of $\sim 0.4$~eV, relative to 84.1~eV in pristine Au$_{0.4}$Pt$_{0.6}$ NCs
(Fig.~S12 and Table S3). Similarly, positive shifts also appeared in Pt 4f, S 2p, N 1s,
and O 1s of the grafted Au$_{0.4}$Pt$_{0.6}$-CNT compared with pristine Au$_{0.4}$Pt$_{0.6}$ NCs
(Fig.~S11-12). These global shifts likely arise from sample charging during the XPS measurement
and are not indicative of chemical changes in the NCs. To further elucidate the structure of the
NCs on grafted Au$_{0.4}$Pt$_{0.6}$-CNT, the XPS Au 4f$_{7/2}$ and Pt 4f$_{7/2}$ spectra of
grafted pure Au-CNT and Pt-CNT, respectively, were both measured. As shown in Fig.~S13,
the Au 4f$_{7/2}$ spectrum of the grafted pure Au-CNT can be deconvoluted into two peaks at
84.5~eV (Au(I)) and 84.0~eV (Au(0)), indicating that the NCs in the grafted pure Au-CNT
samples also have an Au(I) core-Au(0) shell structure similar to that of the pristine Au
NCs~\cite{luo_2012}. Moreover, Au(I) and Pt($\delta +$) peaks in Au$_{0.4}$Pt$_{0.6}$-CNT
again reveal the formation of Au(I)/Pt($\delta +$)-thiolate complexes as the shell of the
nanoclusters in Au$_{0.4}$Pt$_{0.6}$-CNT. Notably, the Au(0) (Fig.~S13a) signal can hardly
be observed in Au$_{0.4}$Pt$_{0.6}$-CNT, whilst an obvious Pt(0) (Fig.~S13b) pattern is
present in Au$_{0.4}$Pt$_{0.6}$-CNT, demonstrating that Pt(0) replaces the Au(0) core after
adding Pt to Au-CNT~\cite{christensen_2012}. Therefore, these results indicate a core-shell
NC structure in the grafted Au$_{0.4}$Pt$_{0.6}$-CNT samples with Pt(0) as the core and a
Au(I)/Pt($\delta +$)-thiolate complex as the shell. 

\subsection{XPS characterization of annealed ligand-free Au$_{0.4}$Pt$_{0.6}$-CNT-A}

The dominant peak in the Pt 4f spectrum of the Au$_{0.4}$Pt$_{0.6}$-CNT-A (\fig{01}h) at 72.0~eV
(Pt($\delta +$)) shows a negative shift of $\sim 0.7$~eV, relative to 72.7~eV (Pt($\delta +$))
in the Au$_{0.4}$Pt$_{0.6}$-CNT spectrum. Similarly, the dominant peak in the Au 4f spectrum
of the Au$_{0.4}$Pt$_{0.6}$-CNT-A sample (\fig{01}h and Table S3) at 83.7~eV (Au(0)) shows a
negative shift of $\sim 0.8$~eV, relative to the peak at 84.5~eV in Au$_{0.4}$Pt$_{0.6}$-CNT;
the peak at 70.9~eV of the Pt 4f$_{7/2}$ spectrum can be assigned to Pt(0)~\cite{imaoka_2017},
suggesting an increased metallic character of the AuPt NCs on CNTs after annealing.
The presence of the Pt(0) peak and the increased intensity of the Pt($\delta +$)
in the alloyed vs the pure Pt sample suggests the existence of both a pure Pt phase and
an AuPt alloyed phase. In combination with the other results showed later in this paper,
we take this as a strong indication that the structure of the AuPt NCs is that of a Pt core
surrounded by an AuPt shell. Besides, the Au$_{0.4}$Pt$_{0.6}$-CNT-A XPS shows a high content
of S-O bonds, with a dominant S 2p$_{3/2}$ peak at 167.91~eV, and a severely diminished
presence of S in thiol groups (Fig.~S11 and Table S2). This is a strong indication of the
breakdown during annealing of the thiol groups initially attached to the AuPt NCs. A significant
amount of FeS$_2$ and very minor AuS/PtS components are also identified in the XPS analysis,
at binding energies of 161.85~eV and 165.83~eV, respectively. Low amounts of PtO and PtO$_2$
were observed in the ligand-free Au$_{0.4}$Pt$_{0.6}$-CNT-A sample due to surface oxidation.

\subsection{XANES characterizations}

To further confirm the oxidation state of Pt in the Au$_{0.4}$Pt$_{0.6}$-CNT-A sample,
XANES spectra (Figs.~\ref{01}i and S14) were also collected. The Pt L$_2$ and Pt L$_3$
edges show that the Au$_{0.4}$Pt$_{0.6}$-CNT and Au$_{0.4}$Pt$_{0.6}$-CNT-A samples display
similar patterns as Pt foil, indicating that the Pt element in all the samples is mostly
in metallic state, confirming Pt(0) in the core of both Au$_{0.4}$Pt$_{0.6}$-CNT and
Au$_{0.4}$Pt$_{0.6}$-CNT-A, which is consistent with the XPS results of Pt 4f spectra
(\fig{01}h). For the Pt L$_2$ edge, the white line is significantly higher for the
ligand-free AuPt (Pt)-CNT-A samples than for the Pt foil, which means there are some empty
Pt d$_{3/2}$ states close to the Fermi level, which may be the oxidized state of Pt. This
is consistent with the XPS results. The (ligand-protected) AuPt-CNT shows a higher white line
than (ligand-free) AuPt-CNT-A, suggesting the dominant features of AuPt-CNT are situated between
those of the Pt foil (Pt(0)) and PtO$_2$ (Pt(IV)).

\subsection{Effect of varying composition on the results}

\begin{figure*}[t]
    \centering
    \includegraphics[width=0.8\linewidth]{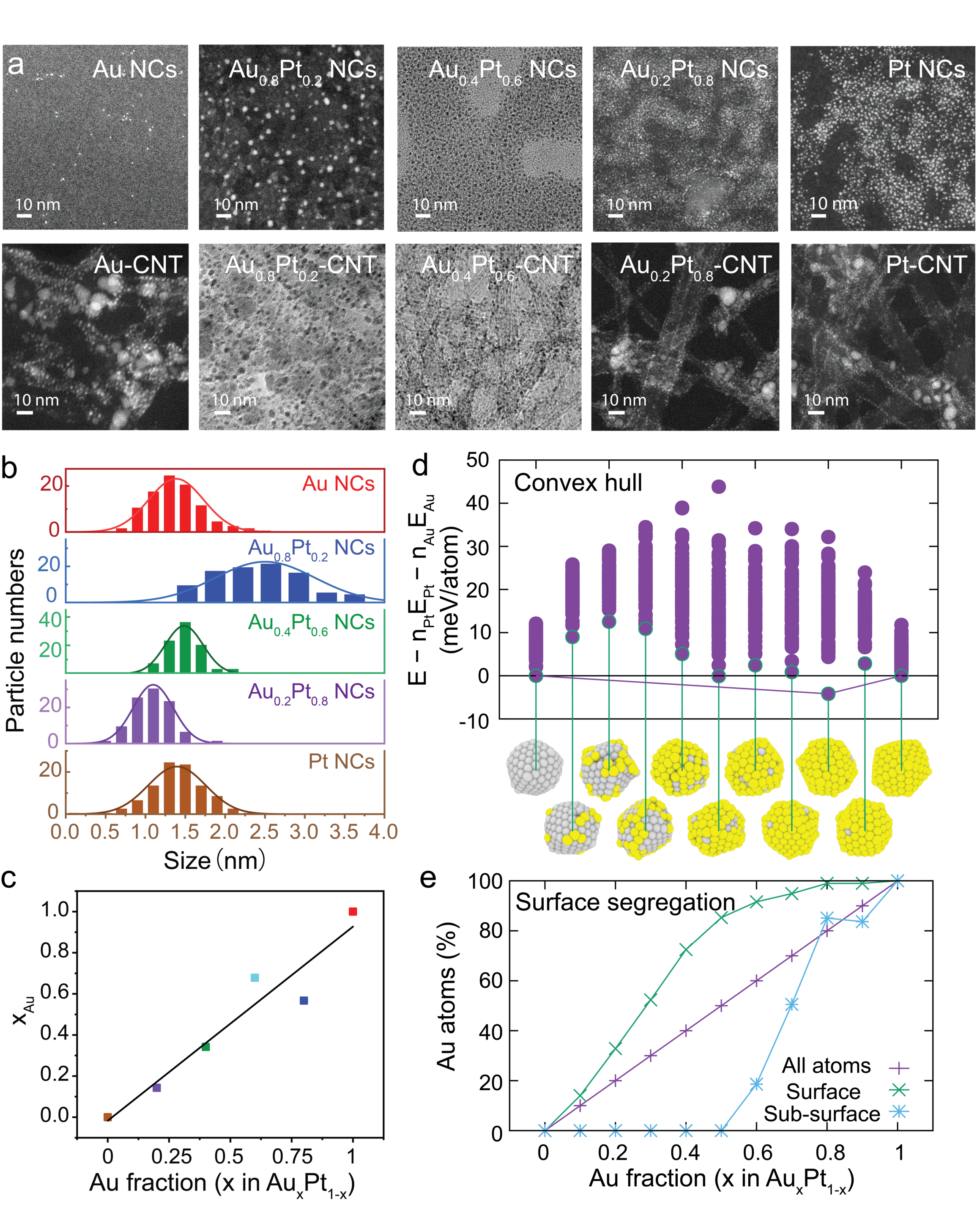}
    \caption{a) STEM BF/DF images of AuPt NCs and AuPt-CNT with various precursor ratios
 in the reaction mixture. AuPt NCs show bright contrast in dark-field view. Vice versa,
 AuPt NCs show dark contrast in bright-field view. b) Corresponding particle-size statistics
    of AuPt NCs. c) Atomic Au fraction as the function of the Au-precursor (HAuCl$_4$)
    concentration present in AuPt-CNT. d) Convex-hull reconstruction of AuPt NCs with
    350 atoms as a function of composition. At each composition, 70 candidate NPs are
    generated and the lowest in energy is chosen to draw the convex hull. These
    low-energy NCs are also depicted below. e) Percentage of Au surface and subsurface
    atoms in the low-energy NCs, as a function of composition.}
    \label{02}
\end{figure*}

We explore different structures of the AuPt-CNT-A samples by varying the atomic Au/Pt
ratios. The nominal compositions of pure (ligand-protected) Au, Au$_{0.8}$Pt$_{0.2}$, Au$_{0.4}$Pt$_{0.6}$,
Au$_{0.2}$Pt$_{0.8}$, and pure Pt NCs as the precursors are achieved by varying the
HAuCl$_4$ to H$_2$PtCl$_6$ ratios in the reaction mixture, respectively.
TEM images of these samples are shown
in \fig{02}a and Figs.~S15-17. The corresponding size distribution of the discrete NCs
is as shown in \fig{02}b, indicating average sizes of $\sim 1.5$~nm for Au, Au$_{0.4}$Pt$_{0.6}$,
Au$_{0.2}$Pt$_{0.8}$, and pure Pt NCs. The size distribution of Au$_{0.8}$Pt$_{0.2}$ NCs peaks
at a slightly larger value, $\sim 2.5$~nm. The NC sizes in the (ligand-protected) Au-CNT,
Au$_{0.4}$Pt$_{0.6}$-CNT, Au$_{0.2}$Pt$_{0.8}$-CNT, the Pt-CNT samples (\fig{02}a and
Figs.~S18-20) are similar to those before grafting. From the inductively coupled plasma
(ICP) analysis (Table~S6), the (ligand-free) AuPt-CNT-A films with an atomic Au fraction
$x_\text{Au}$ = 1, 0.56, 0.68, 0.34, 0.143, and 0 correspond to the nominal compositions of
Au-CNT-A, Au$_{0.8}$Pt$_{0.2}$-CNT-A, Au$_{0.6}$Pt$_{0.4}$-CNT-A, Au$_{0.4}$Pt$_{0.6}$-CNT-A,
Au$_{0.2}$Pt$_{0.8}$-CNT-A and Pt-CNT-A, respectively (\fig{02}c). Recall that the nominal
compositions are proportional to the relative concentration of HAuCl$_4$ in the reaction mixture,
but this may deviate slightly from the ICP measured values. Moreover, the pure Au-CNT-A
and Pt-CNT-A samples (depicted in Figs.~S21-22) show similar morphology to the grafted CNT
films, demonstrating that the samples retain their overall morphology after annealing. 

To investigate the dependence of the electronic structure on the atomic Au/Pt ratios in
the AuPt-CNT-A films, we measured their XPS spectra (Figs.~S23-24). The main peak position
in the Au 4f$_{7/2}$ spectra shifted from 84.06 down to 84.04, 83.84, 83.64, and 83.48~eV
as the Au/Pt ratio was reduced, which indicates the synergistic effect of Au and Pt.
The Au 4f$_{7/2}$ spectrum of the Au$_{0.4}$Pt$_{0.6}$-CNT-A film negatively shifted by about
$-0.42$~eV compared with Au-CNT-A, whereas the Pt 4f$_{7/2}$ spectrum of the same sample
shifted positively by about $+0.24$~eV compared with Pt-CNT-A, revealing the nature of
the electronic interaction between Au and Pt in Au$_{0.4}$Pt$_{0.6}$-CNT-A, with Au being
a slightly more electronegative species than Pt. Additionally, by normalizing the relative
sensitivity factors of the Pt 4f$_{7/2}$ spectra, the average composition (Table~S7) indicates
that the proportion of Pt(0)/Pt($\delta +$) increases with decreasing Au:Pt atomic ratio
for the annealed samples. It is also clear that the diminishing binding energy of Pt(0)
as the Au atomic ratio increases suggests that the Au/Pt atomic ratio can be used to
modulate the electronic structure of the NCs. 

\subsection{Simulated-annealing molecular dynamics}

We carried out a series of atomistic simulations to understand the structure and
electrocatalytic activity of the NCs. The simulations are carried out with a purposely
trained machine learning interatomic potential (MLP)~\cite{deringer_2019} for the Au-Pt-H
system. We start by generating structural models of the ligand-free AuPt NCs. According to the
size-distribution estimates from experiment, the ligand-free NCs containing circa 350 metal atoms are
the most representative, and so we focus on this size range. We use a high-temperature
simulated-annealing procedure to survey 11 compositions with different Au fraction
at 7 different annealing temperatures, and 10 structures per combination
(for a total of 770 ligand-free NCs)~\cite{kloppenburg_2023,jana_2023}.

The end structures are obtained via a sequence of three steps: high-temperature annealing,
quench to low temperature and optimization to the nearest local minimum of the potential
energy surface. The library of structures, as well as the MLP, have been made available via
Zenodo~\cite{kloppenburg_2024,kloppenburg_2024b}. With the lowest-energy ligand-free NCs per
composition, we reconstruct the convex hull of stability for the AuPt system, shown in
\fig{02}d. In the plot, the formation energies of the alloy NCs are computed with respect
to a reference formation energy linearly interpolated from the pure Pt and Au values.
The convex hull is defined by the ``simplest'' curve that contains all data points from
below, going from $x = 0$ to $x = 1$. We see from the small energy differences that
stable structures can be obtained for all values of Au fraction $x$, with a slight reduction
in relative stability at $x \sim 0.2$ of up to 10~meV/atom, which is still significantly
lower than the thermal kinetic energy per atom available at room temperature ($\sim 39$~meV).
We further note that this is a potential energy analysis, and that the alloys would be
further stabilized by the entropic contribution if looking at the Gibbs free energy of
formation instead. 

At every composition we select the lowest-energy NC for further structural analysis,
and to carry out the hydrogenation simulations. A very obvious structural feature of these
NCs is the strong segregation of Au atoms towards the surface. \fig{02}e shows the percentage
of Au atoms among the surface and subsurface atoms, determined with a stochastic
probe-sphere algorithm~\cite{ase_tools}. At all compositions, a strong increase in the
amount of Au atoms at the surface is observed, compared to the amount expected from the
nominal NC composition. At the same time, a very strong depletion of subsurface Au atoms
takes place. This means that, even for NCs as small as these, where as many as 200 atoms
out of 350 are at the surface, the NCs are almost entirely covered by Au atoms at and
above 60~\% Au content. This confirms the core/shell structure of the AuPt NCs after
annealing observed experimentally. In the NCs with low Au fraction there is a clear
prediction for a Pt(core)/AuPt(shell) structure, in accordance with our experimental
XPS analysis of ligand-free AuPt-CNT-A.

\subsection{Electrocatalytic performance of AuPt-CNT-A for HER}

\begin{figure*}[t]
    \centering
    \includegraphics[width=1\linewidth]{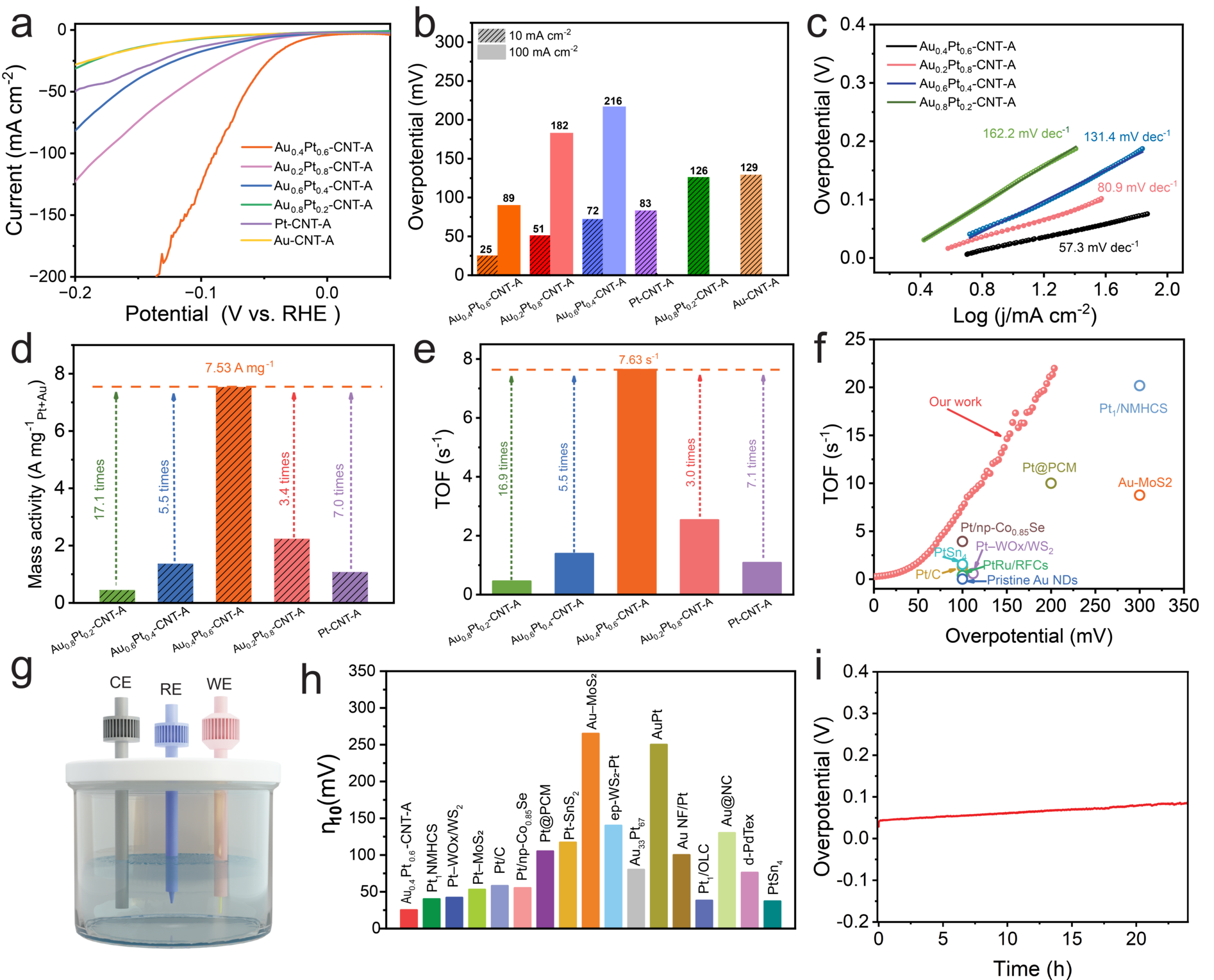}
    \caption{a) Linear sweep voltammetry (LSV) plots of Au-CNT-A, Au$_{0.8}$Pt$_{0.2}$-CNT-A,
    Au$_{0.6}$Pt$_{0.4}$-CNT-A, Au$_{0.4}$Pt$_{0.6}$-CNT-A, Au$_{0.2}$Pt$_{0.8}$-CNT-A,
    and Pt-CNT-A. b) Overpotentials of catalysts at 10~mA~cm$^{-2}$ and 100~mA~cm$^{-2}$.
    c) Tafel plots of Au$_{0.4}$Pt$_{0.6}$-CNT-A and the control samples. d) Mass activity
    at the overpotential of 100~mV in comparison with those of Au$_{0.8}$Pt$_{0.2}$-CNT-A,
    Au$_{0.6}$Pt$_{0.4}$-CNT-A, Au$_{0.2}$Pt$_{0.8}$-CNT-A. e) The turnover frequency (TOF)
    at the overpotential of 100~mV of Au$_{0.8}$Pt$_{0.2}$-CNT-A, Au$_{0.6}$Pt$_{0.4}$-CNT-A,
    Au$_{0.4}$Pt$_{0.6}$-CNT-A, Au$_{0.2}$Pt$_{0.8}$-CNT-A, and Pt-CNT-A. f) Comparison of TOF
    of Au$_{0.8}$Pt$_{0.2}$-CNT-A with recent research. g) Schematic illustration of hydrogen
    evolution reaction. h) Comparison of the overpotential of Au$_{0.8}$Pt$_{0.2}$-CNT-A
    with recent research. i) Chronoamperometric measurement of Au$_{0.4}$Pt$_{0.6}$-CNT-A.
    All tests are performed in N$_2$-saturated 1.0~M H$_2$SO$_4$ solutions.}
    \label{03}
\end{figure*}

Figs.~\ref{03}a and g show the polarization curves of the samples at different
Au/Pt ratios and the schematic illustration of the electrochemical cell setup in a
0.5~M H$_2$SO$_4$ electrolyte, respectively. The pristine CNTs and sulfated CNTs are
inactive for HER, indicating that the Fe catalyst plays a negligible role in HER
(Fig.~S28). Interestingly, the highest catalytic activity is shown by
Au$_{0.4}$Pt$_{0.6}$-CNT-A, exhibiting low overpotentials of 25 and 89~mV at the cathodic
current densities of 10 and 100~mA~cm$^{-2}$ (Figs.~\ref{03}a-b), respectively, which is
comparable to, or even smaller than, those of many reported noble-metal catalysts
(\fig{03}h and
Table S8).
It is superior to both Pt-CNT-A or Au-CNT-A, revealing the synergistic enhancement of
the catalytic activity between Au and Pt in HER. Furthermore, to investigate how the catalytic
activity is affected by different ratios of Au/Pt, the polarization curves were also performed on Au$_{0.8}$Pt$_{0.2}$-CNT-A, Au$_{0.6}$Pt$_{0.4}$-CNT-A, and Au$_{0.2}$Pt$_{0.8}$-CNT-A (\fig{03}a).
As shown in \fig{03}b, the overpotentials required to drive the Au$_{0.8}$Pt$_{0.2}$-CNT-A
cathodic current densities of 10 and 100~mA~cm$^{-2}$ are 126 and 281~mV, respectively;
for Au$_{0.6}$Pt$_{0.4}$-CNT-A they are 72 and 216~mV; and for Au$_{0.2}$Pt$_{0.8}$-CNT-A
they are 51 and 182~mV. Therefore, among the available samples, Au$_{0.4}$Pt$_{0.6}$-CNT-A
shows the highest activity, revealing the optimal concentration of Au to enhance the HER
catalytic performance. Therefore, the high intrinsic activity of Pt can be further increased
by tuning the geometric and electronic structure of AuPt NCs via surface segregation of the
Au atoms, which is confirmed by XPS and XANES analysis and MD simulations (Figs.~\ref{01}g-h,
and Figs.~\ref{02}d-e). The activity of Pt atoms towards H adsorption increases
(by lowering the overpotential) when these are surrounded by Au atoms at the NC facets.
A full mechanistic picture of this process is presented below through H adsorption simulations
at variable chemical potential with the MLP. \fig{03}c displays the Tafel plots (log j vs $\eta$)
of the HER and shows that Au$_{0.4}$Pt$_{0.6}$-CNT-A exhibited a small corresponding Tafel
slope of 57.3~mV~dec$^{-1}$, suggesting that the HER on this catalyst is also dominated by
the Tafel mechanism (H$_\text{ads}$ + H$_\text{ads}$ $\rightarrow$ H$_2 \uparrow$). In contrast,
the Au$_{0.8}$Pt$_{0.2}$-CNT-A, Au$_{0.6}$Pt$_{0.4}$-CNT-A, and Au$_{0.2}$Pt$_{0.8}$-CNT-A
samples have Tafel slopes of 162.2, 131.4, and 80.7~mV~dec$^{-1}$, respectively. This confirms
that Au$_{0.4}$Pt$_{0.6}$-CNT-A has a much faster HER kinetics than NC with other Au/Pt ratios.
The Tafel slopes of the CNTs (figure S28) are 143.2~mV~dec$^{-1}$, 
indicating the slow kinetics of CNT for HER without nanoclusters. The electrochemical
impedance spectroscopy (EIS) analysis further demonstrates the enhanced HER kinetics in the
0.5~M H$_2$SO$_4$ electrolyte (Figs.~S25-26). The charge transfer resistance ($R_\text{ct}$)
is listed in Table~S9. Au$_{0.4}$Pt$_{0.6}$-CNT-A shows a much lower $R_\text{ct}$ (0.25~$\Omega$)
than those of Au$_{0.2}$Pt$_{0.8}$-CNT-A ($R_\text{ct} = 0.5 \Omega$), Pt-CNT-A (1.14~$\Omega$),
and Au-CNT-A (3.25~$\Omega$), indicating that Au$_{0.4}$Pt$_{0.6}$-CNT-A has an improved electron
transport kinetics compared to Pt-CNT-A and Au-CNT-A.
The enhanced catalytic activity can be attributed to the modified electronic structure of the Pt
surface sites on Au$_{0.4}$Pt$_{0.6}$ NCs and their increased ability to bind hydrogen, as will
be explained in detail below in the context of our atomistic simulations. 

The ECSAs of the Au$_{0.4}$Pt$_{0.6}$-CNT-A and referenced samples are shown in Fig.~S27.
The corresponding electric double-layer capacitance ($C_\text{dl}$) value of
Au$_{0.4}$Pt$_{0.6}$-CNT-A is close to that of other samples, indicating the catalysts
have similar number of active sites. Besides, as shown in \fig{03}d, Au$_{0.4}$Pt$_{0.6}$-CNT-A
has a high mass activity~\cite{zhu_2018} of 7.53~A~mg$^{-1}$ at $\eta = 100$~mV, which is
3.4, 5.5, 7, and 17.1 times that of the Au$_{0.2}$Pt$_{0.8}$-CNT-A, Au$_{0.6}$Pt$_{0.4}$-CNT-A,
Pt-CNT-A, and Au$_{0.8}$Pt$_{0.2}$-CNT-A samples, respectively,
indicating an intrinsic HER activity of Au$_{0.4}$Pt$_{0.6}$-CNT-A higher than that
of AuPt-CNT-A with other Au/Pt ratios. 

As shown in \fig{03}e, the turnover frequency (TOF) of the Au$_{0.4}$Pt$_{0.6}$-CNT-A
sample at $\eta = 100$~mV is 7.63~s$^{-1}$, which outperforms the other bimetallic
samples with different Au/Pt ratios studied here as well as the majority of reported Au-
and Pt-based catalysts from the literature (\fig{03}f and Table~S10), further indicating that the Au atoms could work as an enhancer for the intrinsic activity of Pt in Au$_{0.4}$Pt$_{0.6}$-CNT-A by modulate the electronic structure of the ligand-free NCs on CNT. 

It should be noted that, prior to annealing, the Au$_{0.4}$Pt$_{0.6}$-CNT sample (Fig.~S28)
performed far worse than the ligand-free Au$_{0.4}$Pt$_{0.6}$-CNT-A sample did. A
significantly enhanced catalytic activity can be observed after annealing, demonstrating
that the ligand-removal process plays a crucial role in HER enhancement. At the same time, the
grafted (ligand-protected) Au$_{0.4}$Pt$_{0.6}$-CNT shows a higher $R_\text{ct}$ than annealed
ligand-free Au$_{0.4}$Pt$_{0.6}$-CNT-A, and also shows low ECSAs, suggesting the annealing
process can remove the catalytically inert organic groups on the NC surface, whilst
optimizing the NC bimetallic structure. For comparison, as shown in Fig.~S29, the annealed
Au$_{0.4}$Pt$_{0.6}$ NCs without the support of CNT agglomerated. This suggests the CNT film
promotes the uniform distribution of Au$_{0.4}$Pt$_{0.6}$ NCs and prevents agglomeration,
thus maximizing the exposure of active sites. Besides, as shown in Fig.~S30, the polarization
curves of pure ligand-protected Au$_{0.4}$Pt$_{0.6}$ NCs prior to grafting on the CNTs and annealing show a
high overpotential and low mass activity (0.0376~A~mg$^{-1}$), which indicates a low HER activity
of the ligand-protected Au$_{0.4}$Pt$_{0.6}$ NCs. 

The durability of the Au$_{0.4}$Pt$_{0.6}$-CNT-A sample is examined by a chronoamperometric
measurement (\fig{03}i) at 10~mA~cm$^{-2}$ over 24~h without significant variation in
overpotential, suggesting the Au$_{0.4}$Pt$_{0.6}$-CNT-A film electrode is stable for
HER catalysis in acid solution during a long-term durability test. The Au 4f and Pt 4f
XPS spectra were also measured for this sample post chronoamperometric measurement,
showing slightly oxidized Pt on the surface of the Au$_{0.4}$Pt$_{0.6}$-CNT-A
film (Fig.~S31).  

\subsection{Atomistic hydrogenation simulations}

\begin{figure}[t]
    \centering
    \includegraphics[]{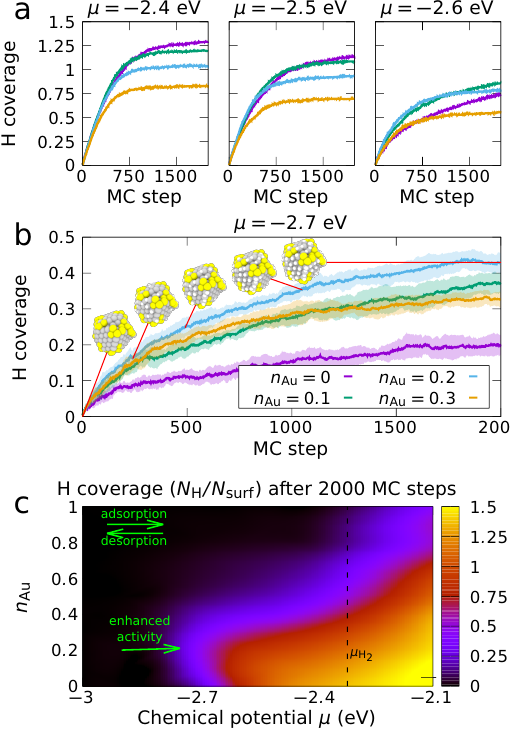}
    \caption{a) Example MCMC runs (each an average over 10 individual MCMC trajectories)
    for 0, 10, 20 and 30~\% Au content (legend in b). b) Detailed MCMC run at $\mu = -2.7$~eV,
    where the shaded areas indicate the statistical error (standard deviation) from averaging
    over the 10 individual trajectories per composition. The inset NC snapshots show the location
    of the adsorbed hydrogens at different stages for a particular trajectory at 20~\%
    Au content. c) Coverage after 2000 MCMC steps as a function of composition and chemical
    potential. Adsorption is favored as one increases the chemical potential and desorption
    is favored in the opposite direction. The chemical potential of H$_2$ (half of the cohesive
    energy of the H$_2$ molecule) is indicated with a vertical line. The region of increased
    activity at around 20~\% Au content is also indicated. All panels use the ratio of adsorbed
    hydrogens $N_\text{H}$ to surface metal atoms $N_\text{surf}$ as the definition of coverage.}
    \label{04}
\end{figure}

To elucidate the hydrogen adsorption mechanism in the ligand-free bimetallic AuPt
NCs and its relation to the enhanced
HER activity observed experimentally, we carry out heuristic Markov-chain Monte Carlo (MCMC)
simulations. In brief, we allow the addition and removal of adsorbed hydrogen atoms until
the following equilibrium is reached, for a given potential $V$: 
\begin{align}
[\text{NC:H}]^{(i)} + \text{H}^* \rightleftharpoons [\text{NC:H}]^{(i)} +
\text{H}^+ (\text{aq/vac}) + e^-_V,
\label{eq:01}
\end{align}
that is, at step $i$ (after sufficiently many steps), the (free) energy difference between
hydrogen adsorption and desorption is zero. For hydrogen desorption, the free energy of the
proton $\mu_{\text{H}^+}$ is given by a chemical reservoir (in aqueous or vacuous environment)
and the free energy of the electron $-eV$ is given by the electrochemical potential $V$.
Within our methodological framework, this is the same as defining an effective chemical
potential for the neutral hydrogen species given by $\mu_\text{H} = \mu_{\text{H}^+} - eV$.
Since $\mu_{\text{H}^+}$ is fixed, we can mimic the effect of tuning $V$ during the
experiment by tuning $\mu_{\text{H}}$ in the simulation. More details about these simulations
are given in the Supplemental Information. Examples of MCMC trajectories for some of the
most interesting cases are given in the top and middle panels of \fig{04}. \fig{04}b also
provides error estimates based on the statistics of 10 independent runs per composition and
chemical potential. At high chemical potential (low electrode potential) the activity towards
hydrogen adsorption goes down as the Au fraction increases, as expected. However, as the
chemical potential is decreased, one can see an enhancement of the hydrogenation activity at
$x \sim 0.2$. \fig{04}b shows a few snapshots along one of the trajectories at $\mu = -2.7$~eV
and $x = 0.2$, where it is clear that H preferentially adsorbs on or between Pt atoms,
including a visible (111)-like facet where the hydrogens adopt the familiar on-top configuration.
Remarkably, the presence of nearby Au atoms works as an enhancer for the reactivity on
the Pt atoms, and indeed H adsorption is more favorable on the ligand-free bimetallic Au$_x$Pt$_{1-x}$ NCs with
$x \sim 0.2$ than on the pure ligand-free metallic Pt NCs at low chemical potential, or high electrode potential,
in excellent qualitative agreement with the trends observed in experiment, although the
nominal composition of the optimum in experiment is higher than in the simulation results.
\fig{04}c provides a more general view of the effect described here.  

More insight into the mechanistic understanding of this enhancement can be obtained by looking
back at \fig{02}d. We see that the relative cohesive energy per atom is lowest in the ligand-free bimetallic AuPt with
a low content of Au. Therefore, within this range of compositions the ligand-free metallic NCs is more susceptible
to lower its cohesive energy via creation of bonds with adsorbed hydrogen. Thus, the Au-mediated
enhancement of the Pt reactivity towards hydrogen adsorption at high electrode potentials
is achieved by creating a slightly less stable NC surface. However, this process only works
at low Au content because a significant number of Pt surface atoms still need to be available
on the surface, given the inability of Au to efficiently bind hydrogen at high electrode
potentials. 

\section{Conclusion}

In summary, we synthesized a series of high-dispersion and ligand-free AuPt bimetallic
nanoclusters-CNT-A complex by grafting nanoclusters on CNT film and subsequently annealing.
We created ligand-free nanoclusters with a unique core (Pt)-shell (AuPt) structure.
Pt atom exposure on the outer surface enhances the electrocatalytic activity for hydrogen
evolution reaction (HER) in aqueous media via a synergistic effect between Pt and Au atoms.
As revealed by molecular dynamics simulations, the core (Pt)-shell (AuPt) structure of the
annealed AuPt NCs at a low Au fraction is achieved through an almost complete Au surface
segregation. At low-enough Au fraction, a sufficient number of catalytically active Pt sites
remain exposed on the surface. The ligand-free Au$_{0.4}$Pt$_{0.6}$ nanoclusters-CNT-A complex shows
the highest HER activity among all nanoclusters, with an overpotential of 27~mV at 10~mA~cm$^{-2}$,
a high mass activity of 7.49~A~mg$^{-1}$ at $\eta = 100$~mV, and a high TOF of 7.63~s$^{-1}$,
outperforming the majority of reported Pt-based catalysts. Markov-chain
Monte Carlo simulations with a custom-made machine-learning potential indicate that the Au
atoms work as an enhancer for the reactivity on Pt atoms, where H adsorption is more favorable
on AuPt at a low Au fraction than on pure Pt. The enhanced HER activity together with the
stability obtained by grafting on the CNT support show great promise for efficient production
of hydrogen. Furthermore, the elucidation of the atomistic mechanism leading to this anomalous
enhancement provides insight into useful design principles to increase the catalytic activity
of Pt-based nanomaterials with HER-inactive atoms at the atomic level~\cite{zheng_2018},
with potential extensions to other materials.

\section*{Supporting Information}

The supporting information contains details of both the experimental characterization
and the atomistic simulations.

\begin{acknowledgments}
We acknowledge funding from the Research Council of Finland (B.~P.: no. 321443 and
352671; O.~I.: Center of Excellence Program in Life-inspired Hybrid Materials,
LIBER, no. 346108; Z.-P.~L: no. 330214 ), China Scholarship Council
(J.~Kang: no. 202008440308, J.S.: no. 202008440311, Z.X: no. 202008440534).
We also thank Ville Liljestr\"om for conducting the WAXS/SAXS measurements. Use of facilities
and technical support were provided by the Aalto University OtaNano-Nanomicroscopy Center,
Bioeconomy Infrastructure, and Raw Materials Research Infrastructures. J.~Kloppenburg
and M.~A.~C. acknowledge financial support from the Research Council of Finland under
grants nos. 329483, 330488 and 347252, as well as computational resources from CSC
(the Finnish IT Center for Science) and Aalto University's Science-IT Project.
We thank Shun Yu for assisting with the beam time proposal. We acknowledge the
MAX IV Laboratory for time in the BALDER beamline under Proposal 20220785.
We thank Konstantin Klementiev for the XANES measurement and valuable discussions.
\end{acknowledgments}

\def\bibsection{}
\section*{References}

\end{document}